\documentclass[journal]{IEEEtran}
\ifCLASSINFOpdf
\else
\usepackage[dvips]{graphicx}
\fi
\usepackage{url}

\usepackage{graphicx}
\usepackage{amssymb,amsmath,amsfonts,amsthm}
\usepackage{mathrsfs}
\usepackage{cite}
\usepackage{array}
\usepackage{tabularx}
\usepackage[table]{xcolor}
\usepackage{color}
\usepackage{mathtools}
\usepackage{subcaption}
\usepackage{mathtools}
\usepackage{tikz,pgf}
\usetikzlibrary{calc,positioning,mindmap,trees,decorations.pathreplacing}
\usepackage{graphicx}
\usepackage{bbm}
\usepackage[utf8]{inputenc}
\usepackage{algorithm}

\newcommand{\norm}[1]{\left\lVert#1\right\rVert}

\begin{document}
	\title{Consensus-Based Distributed Computation of Link-Based Network Metrics}
		
	\author{Zheng~Chen,~\IEEEmembership{Member,~IEEE}, and~Erik~G.~Larsson,~\IEEEmembership{Fellow, IEEE}
	\thanks{This work was supported in part by Excellence Center at Link\"{o}ping - Lund in Information Technology (ELLIIT).}
	\thanks{Z. Chen and E. G. Larsson are with the Department of Electrical Engineering (ISY), Link\"{o}ping University, Link\"{o}ping, Sweden (email: zheng.chen@liu.se, erik.g.larsson@liu.se).}
}
	\maketitle

\begin{abstract}	
Average consensus algorithms have wide applications in distributed computing systems where all the nodes agree on the average value of their initial states by only exchanging information with their local neighbors. In this letter, we look into \textit{link-based} network metrics which are polynomial functions of pair-wise node attributes defined over the links in a network. Different from  node-based average consensus, such link-based metrics depend on both the distribution of node attributes and the underlying network topology. We propose a general algorithm using the weighted average consensus protocol for the distributed computation of link-based network metrics and provide the convergence conditions and convergence rate analysis.
\end{abstract}

\begin{IEEEkeywords}
	Weighted average consensus, distributed computation, total variation, convergence.
\end{IEEEkeywords}
\vspace{-0.3cm}

\IEEEpeerreviewmaketitle
\section{Introduction}
\IEEEPARstart{C}{onsensus} mechanism is the foundation of distributed systems in the absence of centralized management, which is a long-standing research direction with promising applications in diverse fields, such as swarm robotics \cite{swarm}, opinion dynamics in social networks \cite{opinion}, and emerging decentralized machine learning applications \cite{FL-consensus, SO-ML}. Consensus problems refer to the study of reaching an agreement among multiple agents by only  exchanging information with their local neighbors.
One of the fundamental works on consensus problems is \cite{degroot1974}, where DeGroot gave the convergence conditions for a group of individuals to reach consensus on a subjective probability distribution. 
An overview of consensus algorithms and their performance analysis in multi-agent networks was provided in \cite{consensus-cooperation}, where the convergence conditions were summarized for the class of linear consensus algorithms. 

The most common consensus problem is the average consensus, where all the nodes asymptotically reach consensus on the arithmetic mean of their initial states. This is also referred to as the average consensus gossiping, which is often used to model the diffusion of opinions in social networks\cite{diffusion-belief}. 
In the general case, a group of network nodes can reach consensus on the weighted average of their initial states by appropriately injecting weight coefficients in the linear iteration. In \cite{convergence-weighted}, the authors provided a sufficient condition for the convergence of discrete-time weighted average consensus by using the Perron-Frobenius theorem.
Numerous schemes for accelerating the convergence in distributed average consensus can be found in the literature, such as using extrapolation \cite{consensus_extrapolation}, incorporating localized node state prediction \cite{accelerated}, and applying decreasing step size sequence \cite{step_size}.
In addition to computing the arithmetic average of the initial states of network nodes, consensus mechanism can also be used for distributed reconstruction of a common message  \cite{in-network-processing}, and for estimating some graph-related quantities, such as the spectral radius of a graph \cite{spectral-consensus}.

For the average consensus problems, the objective is to reach agreement on the average value of some node-based quantity. Such metrics cannot reflect the graph structure or the correlation between connected nodes in the graph. For example, in \cite{gaitonde2020adversarial}, the authors introduced a metric named the network disagreement level, which is defined as the sum of the squared difference of the opinions between two connected nodes sharing a common edge. To the best of our knowledge, there is no study yet on distributed computation of network metrics defined over the links/edges of a network. 
 
In this work, we use distributed consensus algorithms for decentralized computation of network metrics which are polynomial functions of some \emph{link-based} pair-wise node attributes. Note that our objective is to compute network metrics defined over the links in a network, which depend on not only the node attribute distribution, but also the underlying network topology. This stands in contrast to existing average consensus algorithms commonly used for computing \emph{node-based} average quantities over all the network nodes, which makes our work the first one of its kind. A practical application of our proposed method is to compute link-based smoothness metrics such as the total variation in graph signal processing \cite{connecting,emerging, graph_sp}. Using our proposed algorithm, each node can obtain an estimate of the total variation by only exchanging information with its neighbors. The convergence conditions and convergence rate are also studied, with some open questions identified for further investigation of this problem.

\vspace{-0.2cm}
\section{Preliminaries: Review of Discrete-Time Weighted Average Consensus\cite{convergence-weighted}}
\vspace{-0.1cm}
Consider a network with $N$ nodes, represented by an undirected graph $\mathcal{G}=\{\mathcal{V},\mathcal{E}\}$ with $\mathcal{V}=\{v_1,v_2,\ldots,v_N\}$ being the node set (vertices) and $\mathcal{E}\subseteq \mathcal{V}\times \mathcal{V}$ being the edge set. The adjacency matrix is $A=[a_{ij}]\in \mathbb{R}^{N\times N}$, where $a_{ij}=1$ if nodes $i$ and $j$ are connected, and $a_{ij}=0$ otherwise. For an undirected graph, we have $A^T=A$. For each node $v_i$, the set of its neighbor nodes is defined by $\mathcal{N}_i=\{v_j\in \mathcal{V}, a_{ij}=1\}$. We denote $d_i=\sum_{j=1}^{N}a_{ij}$ as the degree of node $v_i$, which represents the number of nodes that node $v_i$ is connected to. The degree matrix is written as $D=\text{diag}(d_1,d_2,\ldots,d_N)$. The graph Laplacian is $L=D-A$.

The nodes sharing a common edge can communicate with each other and exchange information about their states. Denote by $x_i(0)$ the initial state of node $v_i$. Suppose that we want all the nodes to reach consensus on the weighted average of the initial states $\mathbf{x}(0)=[x_1(0),\ldots,x_N(0)]^T$, i.e.,
\vspace{-0.1cm}
\begin{equation}
\alpha=\frac{\sum_{i=1}^N w_i x_i(0)}{\sum_{i=1}^N w_i},
\label{eq:consensus-value}
\end{equation} 
 where $w_i$ is a positive weight associated with  node $v_i$. The consensus value $\alpha$ can be reached by the following iteration
% \vspace{-0.1cm}
\begin{equation}
x_i(k+1)=x_i(k)+\frac{\epsilon}{w_i}\sum_{j\in\mathcal{N}_i}\left(x_j(k)-x_i(k)\right), \label{eq:weighted-consensus}
 \vspace{-0.1cm}
\end{equation}
where $\epsilon$ is the step size, and $x_i(k)$ is the state of node $v_i$ in iteration $k$.
On matrix form \eqref{eq:weighted-consensus} is written as
\vspace{-0.1cm}
\begin{equation}
\mathbf{x}(k+1)=P_{w} \mathbf{x}(k),
\label{eq:matrix-form}
\vspace{-0.1cm}
\end{equation}
where $P_w=I-\epsilon W^{-1}L$, and $W=\text{diag}(w_1,w_2,\ldots,w_N)$. 

If the graph $G$ is connected, and if $0<\epsilon<\min_{i} (w_i/d_i)$, then $P_w$ is a primitive matrix, which follows from Theorem 8.5.2 in \cite{horn2012matrix}. Then, by using the Perron-Frobenius theorem, one can show that the weighted average value in \eqref{eq:consensus-value} is asymptotically reached by all the network nodes if the iteration in \eqref{eq:weighted-consensus} is applied. A detailed proof is provided in \cite{convergence-weighted}.
When all the weights $\{w_i\}$ are equal, we get the well-known result for the average consensus problem \cite{consensus-cooperation}.

For a network represented by the undirected graph $\mathcal{G}$, we assume that each node $v_i\in\mathcal{V} $ is associated with an attribute $y_i$, which is a real-valued and positive scalar. Existing consensus algorithms such as the iteration in \eqref{eq:matrix-form} allow us to compute the arithmetic mean of the node attributes $\frac{1}{N}\sum_{i=1}^N y_i$, or more generally, $\frac{1}{N}\sum_{i=1}^N f(y_i)$ for any function $f(\cdot)$.

\section{Distributed Computation of Link-Based Network Metrics}
This section contains the main contribution of this paper.

When the connectivity between network nodes is reflected in some network metrics, the following question naturally arises: can we compute the arithmetic average of any polynomial function of pair-wise node attributes $f(y_i,y_j)$ over the links $(v_i, v_j)\in\mathcal{E}$, using node-based consensus protocols?
We start from an example of a link-based network metric, namely the network disagreement \cite{gaitonde2020adversarial}, or the total variation \cite{emerging, connecting}, and then extend our analysis to the general case with any link-based polynomial attribute function.
\vspace{-0.25cm}
\subsection{Total Variation}
\vspace{-0.1cm}
\label{sec:tv}
For a randomly sampled link in the network, we get a pair of random variables denoted by $(y_i, y_j)$. The expected mean-square deviation, or the total variation \cite{emerging, connecting}, is given by
\vspace{-0.5cm}
\begin{align}
T_y=\frac{1}{M}\!\!\sum\limits_{(v_i,v_j)\in\mathcal{E}}\!\!(y_i-y_j)^2 =\frac{1}{2M} \sum\limits_{i=1}^{N}\sum\limits_{j=1}^{N}a_{ij} (y_i-y_j)^2,
\vspace{-0.2cm}
\end{align} 
where $M=|\mathcal{E}|=\frac{1}{2}\sum_{i=1}^{N} d_i$ is the number of links in the network. 
It can be equivalently written as
\vspace{-0.1cm}
\begin{align}
T_y&=\frac{1}{2M} \sum\limits_{i=1}^{N}\sum\limits_{j=1}^{N}a_{ij} (y_i^2+y_j^2-2 y_iy_j)\nonumber\\
&=\frac{1}{M} \left(\sum\limits_{i=1}^{N}d_i y_i^2-\sum\limits_{i=1}^{N}\sum\limits_{j=1}^{N}a_{ij}y_i y_j\right),
\label{eq:ty}
\end{align} 
which follows from $\sum_{i=1}^{N}a_{ij}=d_j$ and $\sum_{j=1}^{N}a_{ij}=d_i$.

The total variation contains one term which is the weighted average of the squared node attributes, and another term that contains the average cross product of the node attributes for each pair of connected nodes. These quantities can be computed by using three distributed consensus algorithms sequentially and aggregating the resulting consensus values. 

\subsubsection{Step 1}
Since $\sum_{i=1}^{N}d_i=2M$, the first  term of \eqref{eq:ty} can be obtained by applying the weighted average consensus algorithm in \eqref{eq:weighted-consensus} with $w_i=d_i$, $x_i(0)=y_i^2$ and $0<\epsilon<1$. Denoting by $\alpha_1$ the consensus value,  we have
\vspace{-0.15cm}
\begin{equation}
\alpha_1=\frac{1}{\sum_{i=1}^{N}d_i} \sum\limits_{i=1}^{N}d_i y_i^2.
\label{eq:alpha1}
\end{equation}
Then the first term of $T_y$ in \eqref{eq:ty} is obtained as
\begin{equation}
\vspace{-0.15cm}
\frac{1}{M} \sum\limits_{i=1}^{N}d_i y_i^2=2\alpha_1.
\end{equation}
 
\subsubsection{Step 2}
The computation of the second term in \eqref{eq:ty} requires two weighted average consensus (WAC) algorithms, namely WAC1 and WAC2, described in what follows.
 
First, every node $v_i\in \mathcal{V}$ collects the attribute values from its neighbors and obtains the following quantity
\vspace{-0.15cm}
\begin{equation}
\vspace{-0.15cm}
w_i=\sum\limits_{j=1}^{N} a_{ij}y_j=\sum\limits_{j\in\mathcal{N}_i}y_j. 
\label{eq:weights}
\end{equation}
Since $y_i$ are positive, the weights $w_i$ are also positive. Given that most real-world graphs are sparse \cite{barabasi2016network}, the computational complexity is for the initial step is $\mathcal{O}(<k>)$, where $<k>$ is the average degree of the nodes.

Then, we apply an algorithm that we refer to as WAC1, which uses the linear iterative equation in \eqref{eq:weighted-consensus} with the initial states as $x_i(0)=y_i$, and the weights $w_i$ given by \eqref{eq:weights}. The step size $\epsilon$ needs to satisfy
\vspace{-0.15cm}
\begin{equation}
0<\epsilon<\Delta_1=\min\limits_{i} \frac{\sum_{j\in\mathcal{N}_i}y_j}{d_i}.
\label{eq:delta1}
\end{equation} 
The value of $\Delta_1$ can be found by applying a minimum consensus algorithm where each node collects the states of its neighbors and updates its own state by the following rule
\begin{equation}
\vspace{-0.15cm}
x_i(k+1)=\min\limits_{j\in\mathcal{N}_i}\{x_i(k), x_j(k)\},
\end{equation}
with the initial states as $
x_i(0)=\sum_{j\in\mathcal{N}_i}y_j/d_i$.

The number of iterations required to reach consensus is the maximum length of the shortest paths from the node with the minimum initial value to all other nodes in the graph, which is upper bounded by the diameter of the graph. For most real-world networks, the diameter of a graph is  $\mathcal{O}(\log N)$, which is called the \textit{small world phenomenon} \cite{barabasi2016network}. 

Provided that \eqref{eq:delta1} holds, WAC1 algorithm converges to the consensus value
\begin{equation}
\alpha_2=\frac{\sum_{i=1}^N \sum_{j=1}^{N} a_{ij}y_j y_i}{\sum_{i=1}^N \sum_{j=1}^{N} a_{ij}y_j}=\frac{\sum_{i=1}^N \sum_{j=1}^{N} a_{ij}y_j y_i}{\sum_{i=1}^N d_i y_i}.
\label{eq:alpha2}
\end{equation} 
Note that the numerator in \eqref{eq:alpha2} is related to the second term in \eqref{eq:ty} which needs to be computed.
   
To obtain the denominator in \eqref{eq:alpha2}, we apply another algorithm that we refer to as WAC2, which uses the iteration in \eqref{eq:weighted-consensus} with the initial states $x_i(0)=y_i$, the weights $w_i=d_i$, and $0<\epsilon<1$. The consensus value will be reached as 
\vspace{-0.2cm}
\begin{equation}
\alpha_3=\frac{1}{2M}\sum\limits_{i=1}^{N}d_i y_i.
\end{equation} 
Then we obtain the denominator in \eqref{eq:alpha2} as
\vspace{-0.2cm}
\begin{equation}
\sum\limits_{i=1}^{N}d_i y_i=2M\alpha_3.
\label{eq:alpha3}
\vspace{-0.1cm}
\end{equation}
\vspace{-0.1cm}
\subsubsection{Consensus Value Aggregation}
Combining \eqref{eq:alpha1}, \eqref{eq:alpha2} and \eqref{eq:alpha3}, the total variation in \eqref{eq:ty} is obtained by
\vspace{-0.1cm}
\begin{equation}
T_y=2\alpha_1 -2\alpha_2\alpha_3.
\label{eq:tv}
\vspace{-0.05cm}
\end{equation}

\vspace{-0.3cm}
\subsection{Extension to Arbitrary Link-Based Attribute Function}
Let $f(y_i,y_j)$ be a polynomial function of the pair-wise node attributes $(y_i,y_j)$ over the link $(v_i,v_j)\in\mathcal{E}$, given as
\vspace{-0.15cm}
\begin{equation}
f(y_i,y_j)=\sum\limits_{l=0}^{J} \sum\limits_{k=0}^{J} c_{lk} (y_i)^l (y_j)^k,
\label{eq:polynomial}
\end{equation}
where $c_{lk}$ is a constant factor and $J$ is the maximum degree of the polynomials.
The goal is to compute
\vspace{-0.1cm}
\begin{equation}
h(\mathcal{G})=\frac{1}{M}\sum\limits_{(v_i,v_j)\in\mathcal{E}}f(y_i,y_j),
\vspace{-0.15cm}
\end{equation} 
which is summed over all the links/edges in the network.
According to the Weierstrass approximation theorem \cite[Theorem 7.26]{rudin1964principles}, every continuous function defined in a closed interval can be approximated by a sequence of polynomial functions.
Therefore, the possible applications of our work are not limited to polynomial-type network metrics. 

Since $f(y_i,y_j)$ has at most $(J+1)^2$ terms, we can compute each term separately. For the term $(l,k)$, we need to compute 
\vspace{-0.5cm}
\begin{align}
h_{l,k}(\mathcal{G})&=\frac{1}{M}\sum\limits_{i=1}^{N} \sum\limits_{j=1}^{N} a_{ij} c_{lk} (y_i)^l (y_j)^k\nonumber\\
&=\frac{c_{lk}}{M}\sum\limits_{i=1}^{N} (y_i)^l \sum\limits_{j=1}^{N} a_{ij}(y_j)^k.
\vspace{-0.1cm}
\end{align}
Following similar steps as in the total variation case, $h_{l,k}(\mathcal{G})$ can be computed as follows:
\begin{enumerate}
\item Every node collects the attribute values from its neighbors and obtains the following quantity 
\vspace{-0.2cm}
\begin{equation}
w_i=\sum\limits_{j=1}^{N} a_{ij}(y_j)^k=\sum\limits_{j\in\mathcal{N}_i}(y_j)^k. 
\label{eq:weights_general}
\end{equation}
\item  Apply the WAC1 algorithm with the initial states as $x_i(0)=(y_i)^l$, the weights $w_i$ from \eqref{eq:weights_general}, and a step size $\epsilon$ that satisfies
$0<\epsilon<\frac{\sum_{j\in\mathcal{N}_i}(y_j)^k}{d_i}$.
The asymptotic consensus value is
\begin{align}
\alpha_{1, lk}&=\frac{\sum_{i=1}^N \sum_{j=1}^{N} a_{ij}(y_j)^k (y_i)^l}{\sum_{i=1}^N \sum_{j=1}^{N} a_{ij}(y_j)^k}\nonumber\\
&=\frac{\sum_{i=1}^N \sum_{j=1}^{N} a_{ij}(y_j)^k (y_i)^l}{\sum_{i=1}^N d_i (y_i)^k}.
\end{align} 
\item Apply the WAC2 algorithm with the initial states $x_i(0)=(y_i)^k$, the weights $w_i=d_i$ and the step size $0<\epsilon<1$. The asymptotic consensus value is
\vspace{-0.2cm}
\begin{equation}
\alpha_{2, lk}=\frac{1}{2M}\sum\limits_{i=1}^{N}d_i (y_i)^k.
\end{equation}
\end{enumerate}
Then we have
\begin{equation}
h_{l,k}(\mathcal{G})=2\alpha_{1,lk}\cdot \alpha_{2, lk} \cdot c_{lk},
\end{equation}
and
\begin{equation}
h(\mathcal{G})=\sum\limits_{l=0}^{J} \sum\limits_{k=0}^{J} h_{l,k}(\mathcal{G}).
\end{equation}

\section{Convergence Speed Analysis}
First, we reformulate the linear iteration in \eqref{eq:matrix-form} in an equivalent form. We can rewrite \eqref{eq:matrix-form} as
\begin{equation}
\mathbf{x}(k+1)-\mathbf{x}(k)=-\epsilon W^{-1}L \mathbf{x}(k).
\label{eq:update_new}
\end{equation}
Consider the following decomposition 
\begin{equation}
W^{-1}L=W^{-\frac{1}{2}}\hat{L}W^{\frac{1}{2}},
\end{equation} 
where $W^{\frac{1}{2}}$ is the symmetric square root of $W$ and $\hat{L}=W^{-\frac{1}{2}}LW^{-\frac{1}{2}}$. Multiplying both sides of \eqref{eq:update_new} with $W^{\frac{1}{2}}$ yields
\begin{equation}
W^{\frac{1}{2}}\mathbf{x}(k+1)-W^{\frac{1}{2}}\mathbf{x}(k)=-\epsilon\hat{L}W^{\frac{1}{2}} \mathbf{x}(k).
\end{equation}
Letting $\tilde{\mathbf{x}}(k)=W^{\frac{1}{2}}\mathbf{x}(k)$, we obtain 
\begin{equation}
\tilde{\mathbf{x}}(k+1)=(I-\epsilon \hat{L})\tilde{\mathbf{x}}(k).
\label{eq:consensus_new}
\end{equation}
Let $\tilde{P}=I-\epsilon \hat{L}$ denote the new weight matrix, which is symmetric and has the real-valued eigenvalues.

The convergence speed of the iteration in \eqref{eq:consensus_new} depends on the eigenvalues of the weight matrix $\tilde{P}$. Let $1=\lambda_1\geq \lambda_2\geq\ldots\geq  \lambda_N$ be the eigenvalues of $\tilde{P}$ in decreasing order.\footnote{For the linear iteration in \eqref{eq:consensus_new} to converge to a non-zero quantity, the largest eigenvalue of $\tilde{P}$ must be one \cite{olshevsky2009}.}
The convergence rate is characterized by the asymptotic convergence factor, defined as  \cite{xiao2004fast}
\begin{equation}
\rho=\sup\limits_{\mathbf{x}(0)\neq \mathbf{x}^{*}}\lim\limits_{k\rightarrow \infty} \left(\frac{\norm{\mathbf{x}(k)-\mathbf{x}^{*}}_2}{\norm{\mathbf{x}(0)-\mathbf{x}^{*}}_2}\right)^{1/k}, \label{eq:conv-rate}
\end{equation}
where $\mathbf{x}^{*}=\lim_{k\rightarrow\infty}\mathbf{x}(k)$.
It is further shown in \cite{xiao2004fast} that $\rho=\max\{|\lambda_2|, |\lambda_N|\}$,
which means that the convergence speed is limited by the second-largest-modulus eigenvalue of $\tilde{P}=I-\epsilon \hat{L}$. The smaller $\rho$ is, the faster the consensus algorithm converges. 

The WAC algorithms have exponential convergence, so the time complexity depends on the convergence rate and the desired accuracy. 
Under a given accuracy factor $\zeta$, the convergence time for the WAC2 algorithm with $w_i=d_i$ satisfies $\mathcal{O}\left(N^3 \log(N/\zeta)\right)$, as provided in \cite{olshevsky2009} and \cite{landau1981bounds}.

For a graph $\mathcal{G}$ with a certain weight matrix $W$, choosing a large $\epsilon$ helps the distributed consensus algorithm to converge faster. In the WAC2 algorithm, the step size condition to guarantee convergence is $0<\epsilon<1$, which is independent of the graph structure. In the WAC1 algorithm, as shown in \eqref{eq:delta1}, the maximum step size $\Delta_1$ is determined both by the node attributes and by the network topology. In \eqref{eq:delta1}, $\sum_{j\in\mathcal{N}_i}y_j/d_i$ is the average attribute of the neighbors of node $v_i$. If one node is connected with a set of neighbors holding very small attribute values, then this node will become the bottleneck for the convergence speed of the WAC1 algorithm.

\section{Simulation Results}
We apply our proposed algorithms to a real-world data set obtained from the Gnutella-\#4 peer-to-peer file sharing network \cite{snapnets}.
We obtain a subset of the network by  randomly sampling  nodes and keeping the largest connected component with $N=1075$ nodes and $M=2550$ edges. The initial states of the nodes are randomly generated by an exponential distribution with mean value $5$. 

\begin{figure}[ht!]
	\centering
	\begin{subfigure}{.5\textwidth}
		\centering
		\includegraphics[width=\linewidth]{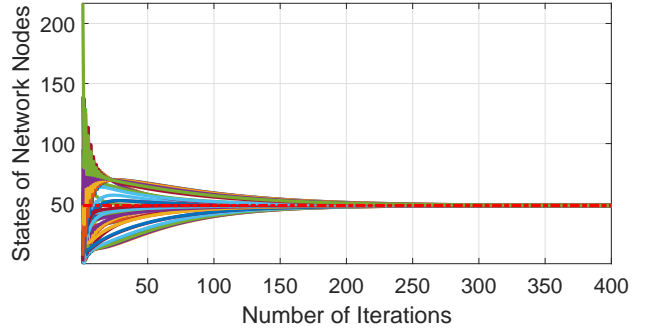}
		\caption{Convergence to consensus value $\alpha_1$.}
	\end{subfigure}
	\begin{subfigure}{.5\textwidth}
		\centering
		\includegraphics[width=\linewidth]{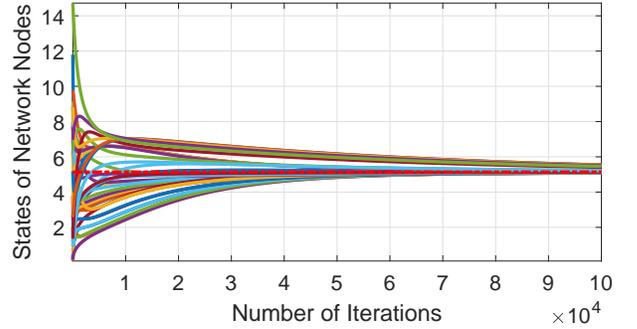}
		\caption{Convergence to consensus value $\alpha_2$.}
	\end{subfigure}
	\begin{subfigure}{.5\textwidth}
		\centering
		\includegraphics[width=\linewidth]{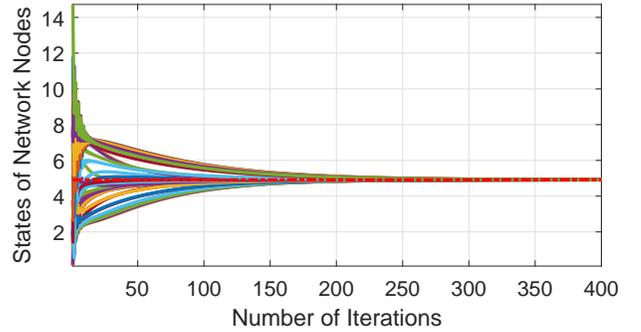}
		\caption{Convergence to consensus value $\alpha_3$.}
	\end{subfigure} 
	\caption{Convergence of the consensus algorithms. }
	\label{fig:consen-convergence}	
\end{figure}

We consider the total variation as the example metric for the simulation. As shown in Section~\ref{sec:tv}, three consensus algorithms are needed to compute the total variation. 
In Fig.~\ref{fig:consen-convergence} we show the convergence of the distributed consensus algorithms. The consensus values $\alpha_1=48.4152$, $\alpha_2=5.138$ and $\alpha_3=4.9122$ give the total variation $T_y=46.3525$ from the relation in \eqref{eq:tv}, which is the same as the true value. The step size $\epsilon$ in each algorithm is chosen to be $0.9 \Delta$, where $\Delta$ is the maximum value of $\epsilon$ that satisfies the convergence condition. Since the step sizes in the algorithms used for $\alpha_1$ and $\alpha_3$ are the same, they have the same convergence speed as we can see from Fig.~\ref{fig:consen-convergence}(a) and (c). However, the convergence time of the algorithm for $\alpha_2$ is much longer, which is related to the distribution of the node attributes.

\begin{figure}[h!]
	\centering
	\includegraphics[width=\linewidth]{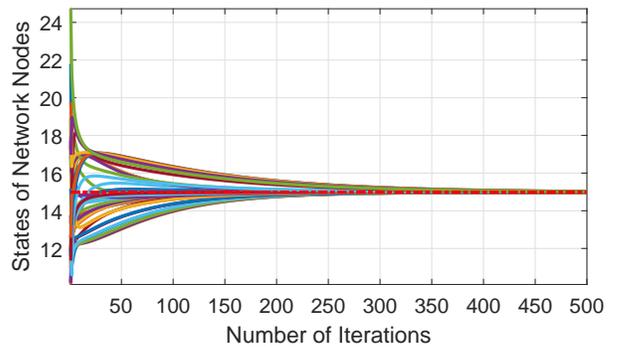}
	\caption{Convergence to consensus value $\alpha_2$ with value shifting.}
	\label{fig:alpha2_alter}
	\vspace{-0.3cm}
\end{figure}

 Note that with the total variation as the metric of interest, its value will not be affected if we add a positive constant to all the node attributes. This technique can give us a larger $\Delta$, which brings the possibility to choose a larger step size $\epsilon$ that might accelerate the convergence of the WAC1 algorithm in Fig.~\ref{fig:consen-convergence}(b). For example, if we add $10$ to all the node attributes, the convergence to the new consensus value $\alpha_2$ is shown in Fig.~\ref{fig:alpha2_alter}.
 However, this value shifting technique is only applicable when the metric is a function of the relative differences between pairs of node attributes.
 Also, while it might improve the convergence speed of WAC1 algorithm, it does not necessarily improve overall convergence.
 More specifically, consider the evaluation of \eqref{eq:tv} after a given finite number $t$ of iterations: $2(\hat\alpha_1^i(t)-\hat\alpha_2^i(t)\hat\alpha_3^i(t))=2(\hat\alpha_1^i(t) - \alpha_2\alpha_3 - \alpha_2\tilde\alpha_3^i(t)-\alpha_3\tilde\alpha_2^i(t)-\tilde\alpha_2^i(t)\tilde\alpha_3^i(t)$, where $\hat\alpha_m^i(t)=\alpha_m+\tilde\alpha_m^i(t)$ is the state value of the node $v_i$ obtained after $t$ iterations and $\tilde\alpha_m^i(t)$ is the deviation from its asymptotic value $\alpha_m$, $\forall m=\{1,2,3\}$. Then, while the shifting technique may reduce $\tilde\alpha_2^i(t)$, it also increases $\alpha_3$, so the total error contribution from $\alpha_3\tilde\alpha_2^i(t)$ may not necessarily decrease for every iteration $t$.
 We leave as an open problem whether a value-shifting or similar technique could be devised that uniformly improves the convergence properties. 
 
 We have also explored the possibility of using weighted edges $a_{i,j}\in[0,1]$ in the adjacency matrix and assigning larger edge weights to the nodes with smaller attributes $y_i$. Simulation results showed that this method does not necessarily improve the convergence performance.

\section{Conclusions}
We studied the application of distributed consensus algorithms for computing network metrics that are defined over the links instead of the nodes in a network. We showed that if all the node attribute values are positive, any arbitrary link-based polynomial function of pair-wise node attributes can be computed by using a sequence of weighted average consensus algorithms.

%\bibliographystyle{IEEEtran}
%\bibliography{ref}
% Generated by IEEEtran.bst, version: 1.12 (2007/01/11)

\end{document}